\documentstyle[12pt,epsf,psfig]{article}
\begin{document}

\begin{center}
{\Large \bf Further application of a semi-microscopic core-particle 
coupling method to the properties of 
$^{155,157}\,$Gd and $^{159}$\,Dy}\\
\end{center}
\begin{center}
{\bf \large
Pavlos Protopapas
\footnote{ pavlos@walet.physics.upenn.edu}
, Abraham Klein
\footnote{ aklein@walet.physics.upenn.edu} ,
 }\\
{\it Department of Physics, University of Pennsylvania, Philadelphia,
PA 19104-6396}  \\
{\bf \large and Niels R. Walet  \footnote{ mccsnrw@afs.mcc.ac.uk}}\\
{\it Department of Physics, UMIST,
P. O. Box 88, 
Machester, M60 1QD, UK}
\end{center}

\date{\today}
\begin{abstract}
In a previous paper a semi-microscopic core-particle coupling method
that includes the conventional strong coupling core-particle model as
a limiting case, was applied to spectra and electromagnetic properties
of several well-deformed odd nuclei. This work, coupled a
large single-particle space to the ground state bands of the
neighboring even cores.  In this paper, we generalize the theory to
include excited bands of the cores, such as beta and gamma bands, and
thereby show that the resulting theory can account for the location and
structure of all bands up to about $1.5$~MeV.
\end{abstract}
\begin{center}
PACS number(s): 21.60.-n, 21.60.Ev, 21.10.-k, 21.10.Re
\end{center}
\section{Introduction}
\label{section:Phonons}
In a previous paper \cite{pavlos:1,pavlos:thesis} we have applied the Kerman-Klein
approach \cite{kk:1,KleinReview,KleinWalet} in the adaptation of
D\"onau and Frauendorf (KKDF) \cite{DF1,fp:11,fp:12} to study
rotational bands of odd deformed nuclei, specifically for the nuclei
$^{157,159}$Gd and $^{159}$Dy. This was done by coupling a
phenomenological rigid rotor, which was described by the
Bohr-Mottelson model, to a single particle. In these applications,
only the ground-state band of the cores was included.  The systems
were described by the conventional monopole pairing plus
quadrupole-quadrupole effective Hamiltonian.

Though the model explored by us is technically more difficult to
implement than the standard particle-rotor models, the results we
found were sufficiently satisfactory that we have been encouraged to
develop a more elaborate version of our work in order to account for
remaining discrepancies. Most striking of the successes is that without
including any {\em ad-hoc} Coriolis attenuation factors we were able
to reproduce the experimentally observed energy levels and
electromagnetic transitions of the lowest bands. Nevertheless, 
there was a shortcoming
for all the applications tried in the
previous work, in that for every nucleus one or more observed bands
at about 1~MeV or higher than the ground state band was missing from
the theoretical results.  It is apparent that these discrepancies,
which we attempt to correct in the present paper, arise from the failure to
include appropriate excited bands of the even core.

Another source of concern about our previous work, the simplified
nature of the Hamiltonian, will not be investigated in the present
work. We remark in passing, however, that some preliminary studies
including quadrupole pairing and hexadecapole interactions have been
carried out for the cases considered here. We found that these
interactions have a small effect on our results if we restrict their
strength to be reasonably smaller ( an order of magnitude smaller)
than the leading interactions (quadrupole-quadrupole and
monopole-pairing).

We return to the main thesis of this paper.  Looking at the
experimentally observed spectra of the even cores for the nuclei under
consideration we can see that $\beta$, $\gamma$ and other higher bands
occur at low energy ($\simeq 1 \rm{MeV}$).  The aim of our treatment
will be to include those excited bands (all highly collective as far
as intra-band transitions are concerned) that are observed to have
non-negligible transition rates to the ground state band.

The general formalism has been described fully in our previous work,
where it was specialized afterwards to the case that the core was represented
only by the ground state band. Therefore, in the following, we shall
consider in detail only those formulas which require generalization
compared to the previous application. 

In the following,  Sec.~\ref{sec:Pho_phe} will be devoted to the
phenomenology of the even cores and to the phenomenological model used to
fit the experimental results.  In Sec.~\ref{sec:Pho_cal} we shall
develop the extension of the KKDF model needed for the inclusion of the
excited bands  and  describe the results of the calculations for
$^{155,157}$Gd, $^{159}$Dy. Finally the last section, \ref{sec:Pho_con}
will contain our concluding remarks.

\section{Phenomenology of the even cores}
\label{sec:Pho_phe}
In the previous work we ignored all excited bands and thus inter-band
transitions were absent from the model. Experimental results justify
this assumption to lowest order, since inter-band BE(2) values, in the
nuclei in which we are interested, are two orders of magnitude smaller
than the intra-band transitions, i.e. quadrupole transitions matrix
elements are an order of magnitude smaller. This assumption worked well
for the low-lying levels (less than $1$MeV excitation).  Nevertheless,
the presence of bands not described by the previous work impel us to
include the effects of excited bands.

\begin{figure}[htbp]
  \begin{center}
    \leavevmode
   \centerline{\hbox{\psfig{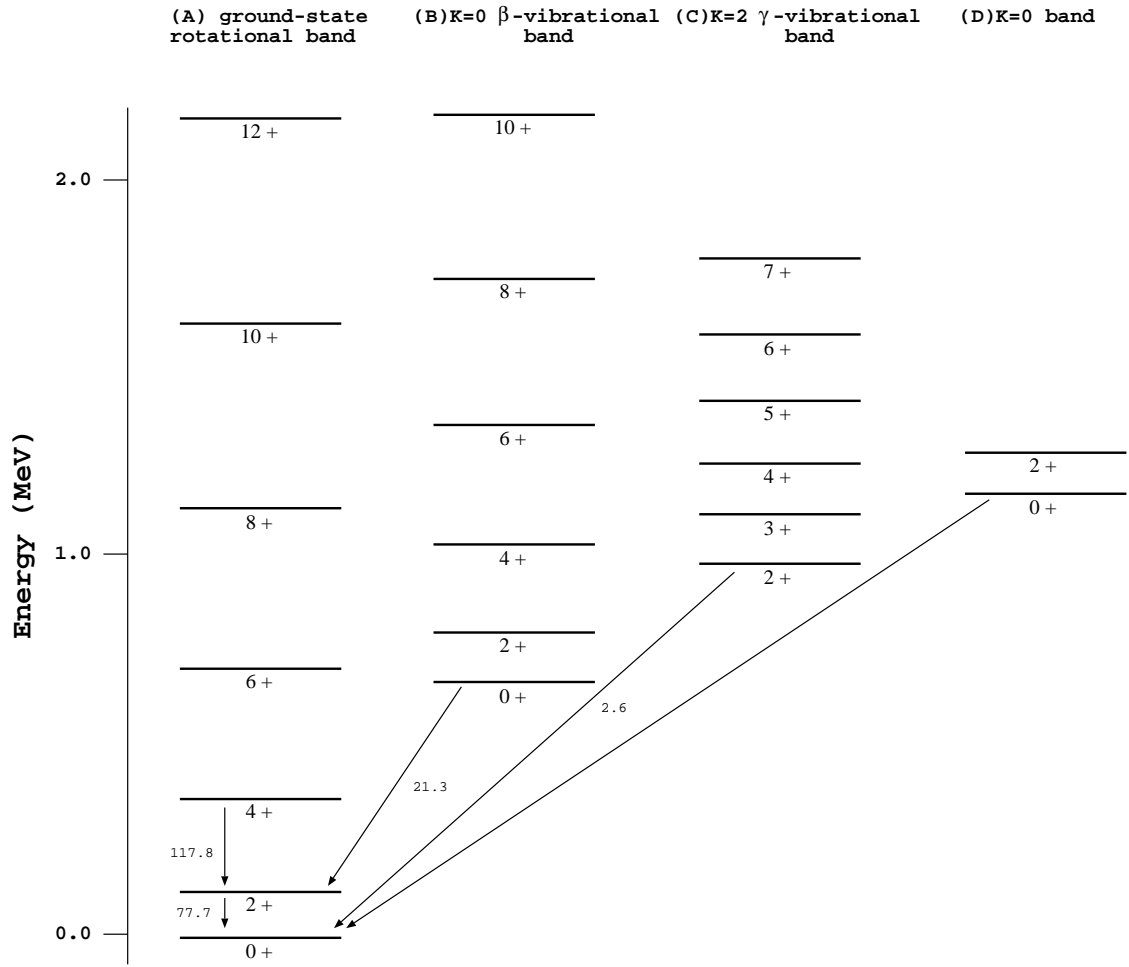}} }
  \end{center}
 \caption{  \label{fig:Gd154} Energy levels and some BE(2) values for $^{154}$Gd. The
BE(2) values are given in $\left[ 10^{-2}(eb)^{2} \right]$. Experimental values for the 
BE(2)'s are taken from Ref.~\protect{\cite{Gupta1}} }
\end{figure}
\begin{figure}[htbp]
  \begin{center}
    \leavevmode
   \centerline{\hbox{\psfig{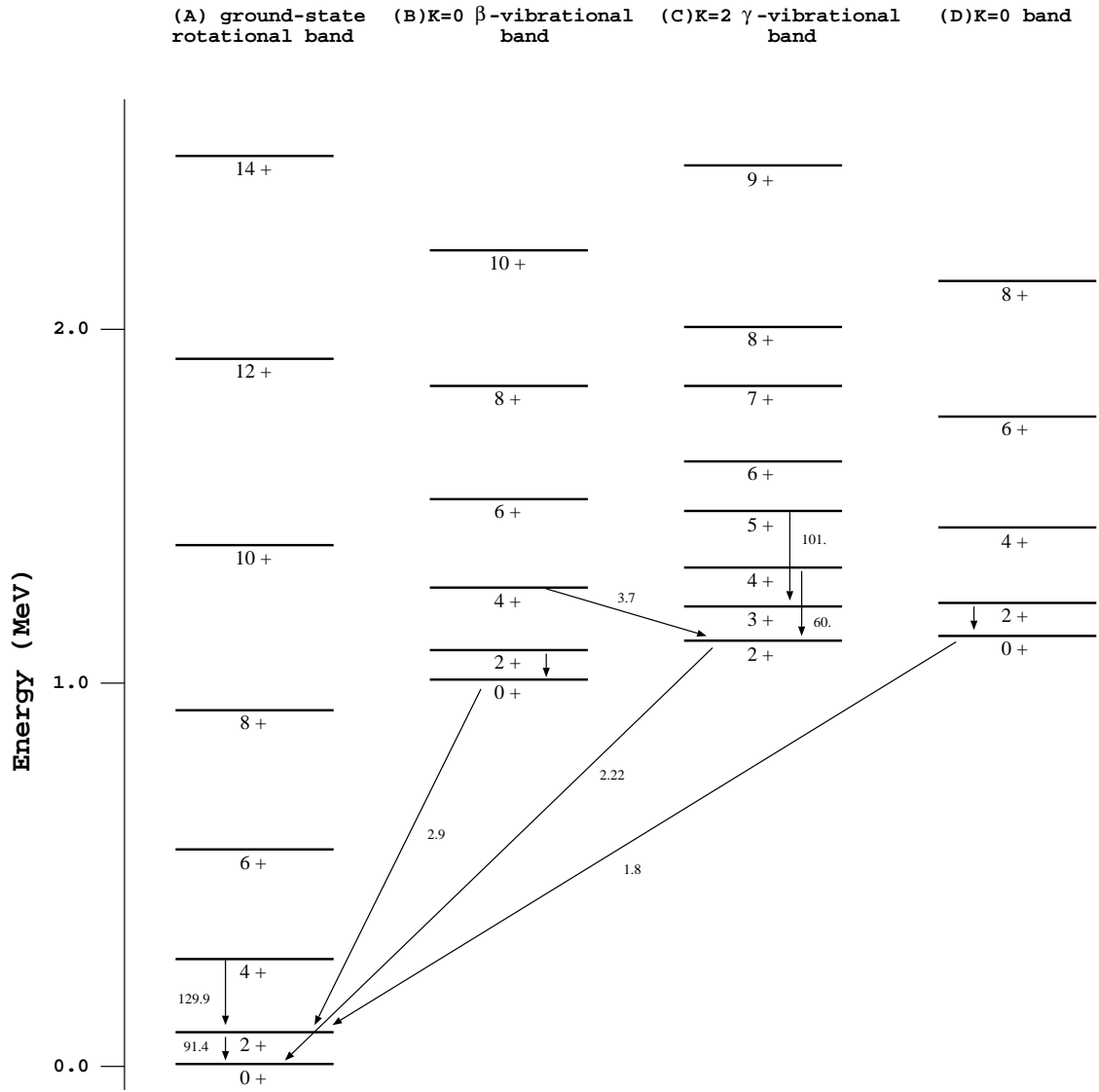}} }
  \end{center}
  \caption{ \label{fig:Gd156} 
Energy levels and some BE(2) values in $\left[10^{-2} (eb)^{2} \right]$ for
$^{156}$Gd. Experimental values for the 
BE(2)'s are taken from Ref.~\protect{\cite{Gupta1}} }
\end{figure}
\begin{figure}[htbp]
  \begin{center}
    \leavevmode
   \centerline{\hbox{\psfig{figure=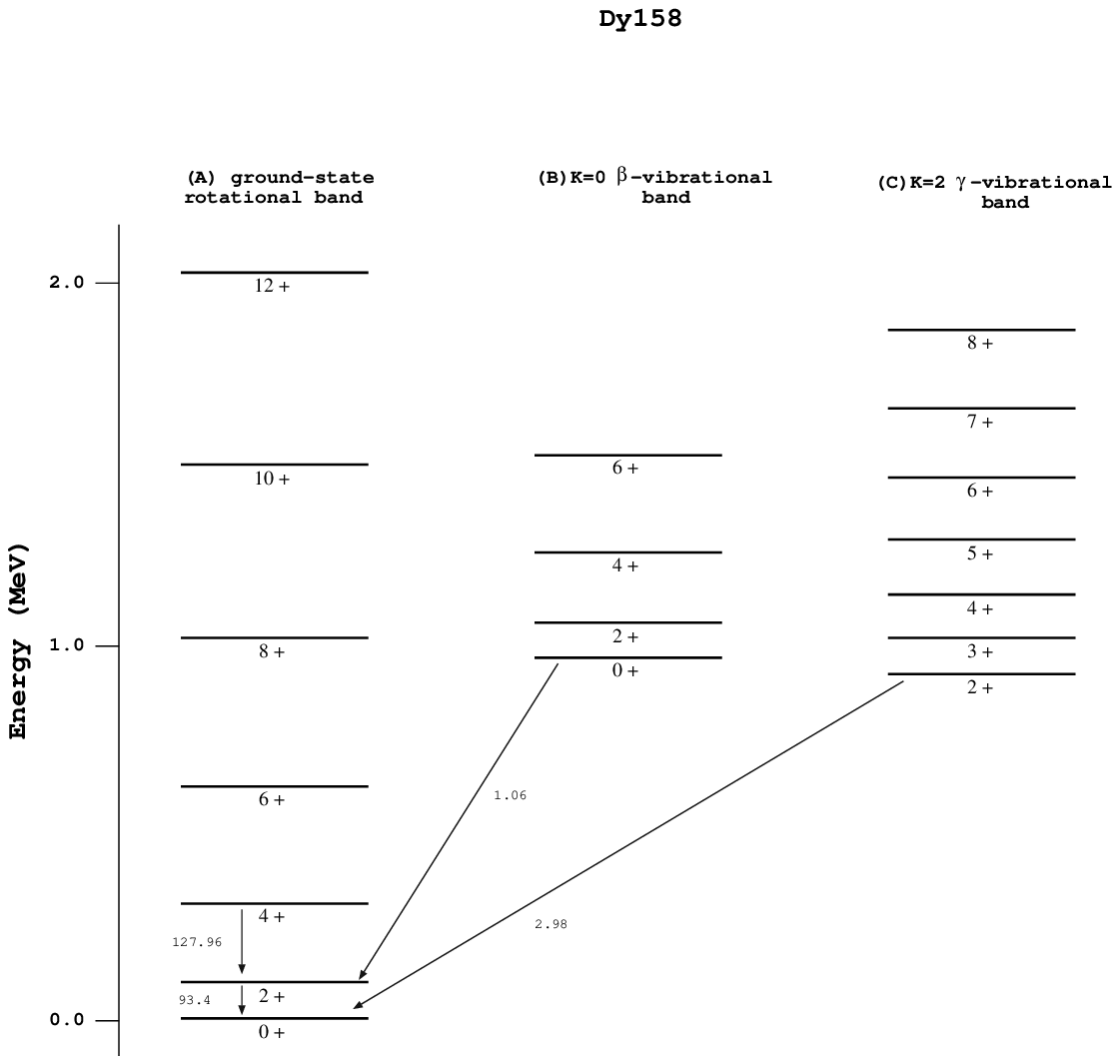,width=15.0cm}} }
  \end{center}
  \caption{ \label{fig:Dy158}
 Energy levels and some BE(2) values in $\left[10^{-2} (eb)^{2} \right]$ for
$^{158}$Dy. Experimental values of BE(2)'s are taken from Ref.~\protect{\cite{Gupta2} }}
\end{figure}

In Figs. \ref{fig:Gd154}, \ref{fig:Gd156} and \ref{fig:Dy158} we
display energy spectra together with observed BE(2) values for intra
and inter-band transitions of cores used in the present study.  As can
be seen from the figures all even nuclei have excited bands at about
1~MeV and the inter-band transitions are small but not zero.  For
example, in the case of $^{156}$Gd we include the $\beta $, $\gamma $
and a third excited band at about 1~MeV above than the ground state
band. As previously stated the values of the inter-band BE(2)
transitions are of the order of 100 smaller than the values of the
corresponding intra-band transitions.  This is a pattern we see in all
the neighboring nuclei. The facts that these transitions are not zero
and the fact that these bands occur at relatively low energies 
justify their inclusion in the calculations.

To perform the calculations for the odd nuclei the excitation energies
and the quadrupole operator matrix elements between any two states of
the even neighbors that belong to the ground-state band or to one of
the excited bands have to be either known from experiment or
calculated from a phenomenological model.  Since there are not
enough experimental values to cover all our needs, we have to use
a phenomenological description to calculate the transition matrix
elements and the excitation energies not available experimentally, i.e.,
we use the phenomenology only to augment experimetal information.

\begin{figure}[htbp]
  \begin{center}
    \leavevmode
   \centerline{\hbox{\psfig{figure=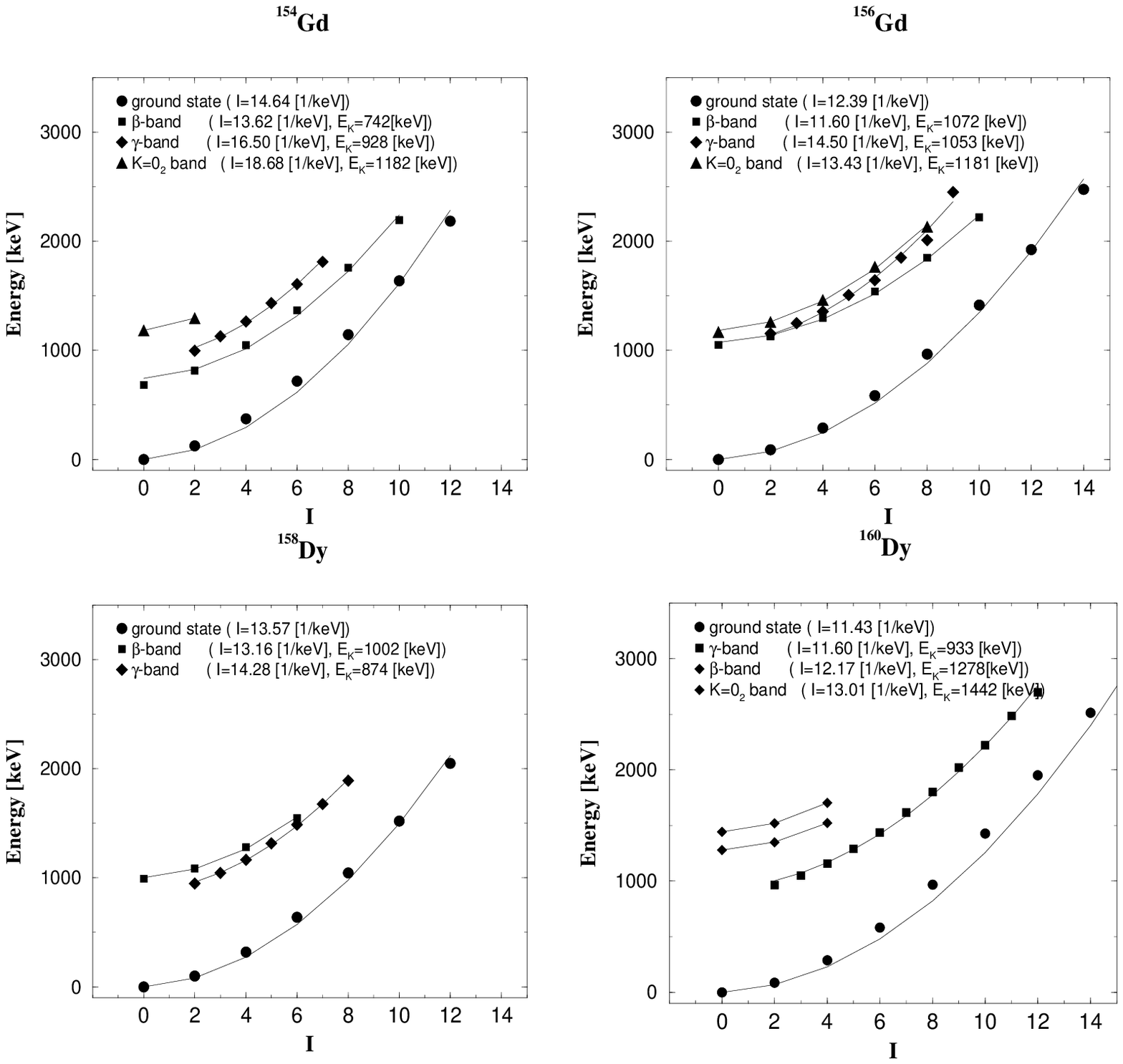,width=15.0cm}} }
  \end{center}
  \caption{  \label{fig:Nuclei_E} 
Experimental and fitted energy levels of $^{154}$Gd, $^{156}$Gd,
$^{158}$Dy and $^{160}$Dy.  Experimental values are shown in filled
symbols and the solid lines represent the theoretical values.  The
band-head energies and the values of the moments of inertia of the
corresponding bands are also given for each band.}
\end{figure}

For the excitation energies we found it sufficient to use the simple formula,
\begin{equation} 
\label{eq:Rigid_rotor_ext_energy}
\omega_{IK} = E_{K} + \frac{ \hbar^{2}}{ 2 {\cal I}_{K}} ( I(I+1)),
\end{equation}
where $E_{K}$ is the band-head energy and ${\cal I}_{K}$ is the moment
of inertia of the given band. In  Fig.~\ref{fig:Nuclei_E} we
show the results of fitting to this formula. We see that this simple
formula is sufficient for the reproduction of the experimental
results, provided we adjust ${\cal I}_{K}$, the individual values
differing from each other by up to $20\%$.

\begin{figure}[htbp]
  \begin{center}
    \leavevmode
   \centerline{\hbox{\psfig{figure=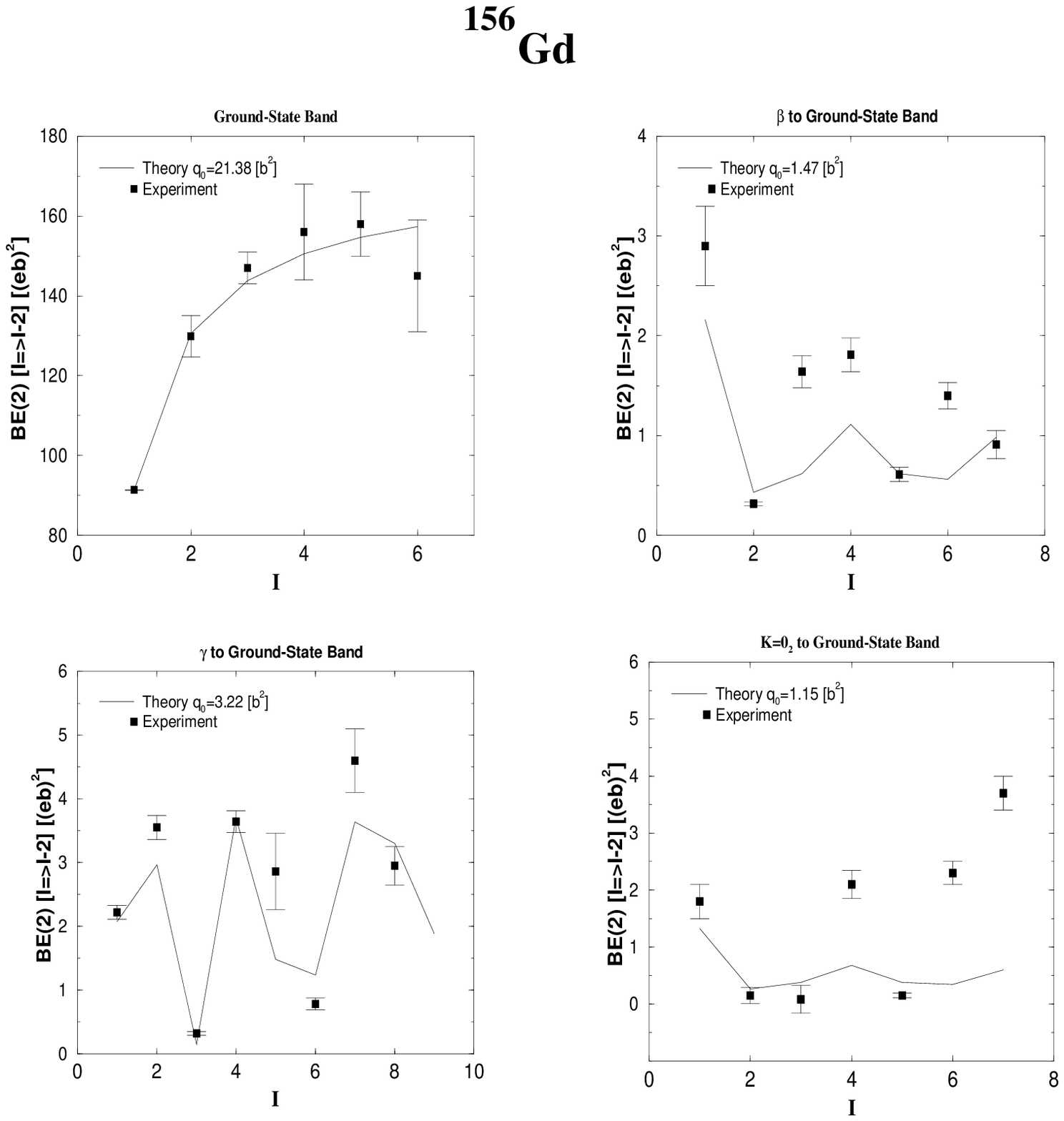,width=15.0cm}} }
  \end{center}
 \caption{   \label{fig:Nuclei_be2}
BE(2) values for $^{156}$Gd. The experimental values are shown as
filled squares and the theoretical predictions are the solid lines. For
every intra or inter-band transitions we give the fitted intrinsic
quadrupole moment $q_{0}$. }
\end{figure}

For the transitional matrix elements we again use the phenomenological
description given by the geometrical model of Bohr and Mottelson,
applicable either to intra or inter-band transitions,

\begin{equation} 
\label{eq:Extended_rigid_rotor_be2}
 \left< IK \| Q \| I'K' \right> =\sigma_{K'}
   q_{0}^{\rm{band1}\rightarrow \rm{band2}} \sqrt{ (2I+1)(2I'+1)}
   \left(
\begin{array}{ccc}
{\scriptstyle I }&{\scriptstyle 2} &{\scriptstyle I'}
\\ 
{\scriptstyle K }&{\scriptstyle -(K-K')} &{\scriptstyle K'}
\end{array} \right),
\end{equation}
where $\sigma_{K'}$ takes the value 1 if $K'=0$ and $\sqrt{2}$ if
$K'\ne 0$.  In Fig.~\ref{fig:Nuclei_be2} we show results of fitting
using the above equation to the experimental BE(2) values for
$^{156}$Gd. Similar fits were performed for all neighboring
nuclei. It is clear that the agreement is far from ideal. 
We emphasize however that experimental values, whether for energy or
matrix elements were used whenever available, and the phenomenological
values were utilized only in the absence of the former (with some
smoothing enforced at the boundary between known and unknown values).

\section{Calculations}
\label{sec:Pho_cal}
Here we present the method and results for energy states in
$^{155}$Gd, $^{157}$Gd and $^{159}$Dy, obtained by coupling a large
single-particle space (single-particle states from 5 major shells were
included) to the ground and excited bands of the appropriate
neighboring even nuclei, using the average description of the latter
implied in the number non-conserving approximation.  As stated above,
the underlying theory, equations of motion, etc, are the same as that
described in detail in our previous work, and therefore will not be
repeated here.  As also noted, all required excitation energies or
transitions (quadrupole transitions) of the even cores were either
calculated or taken from experiment. The matrix elements of the even
cores were expanded to include transitions outside the ground-state
band according to the formula for the core-particle quadrupole interaction, $\Gamma$,
 \begin{eqnarray}
   \label{eq:Gamma_Ph}
     \Gamma_{aIK,cI'K'} &=& -\sigma_{K'}~ \kappa ~ q_{0}^{\rm{band1} 
 \rightarrow \rm{band2}} (-)^{j_{c}+I+J} \nonumber \\
& & \times
 {\setlength{\arraycolsep}{2pt} \left\{\begin{array}{ccc}
 {\scriptstyle j_{a}}&{\scriptstyle {j_{c}}}&{\scriptstyle 2}\\
 {\scriptstyle I'}&{\scriptstyle I}&{\scriptstyle J}
\end{array}\right\}}
               \sqrt{(2I+1)(2I'+1)}
{\setlength{\arraycolsep}{2pt}\left(\begin{array}{ccc} {\scriptstyle
I}&{\scriptstyle 2}&{\scriptstyle I'}\\ {\scriptstyle K}&{\scriptstyle
-(K+K')}&{\scriptstyle K'}\end{array}\right)} q_{ac} ,
\end{eqnarray}
where $\kappa$ is the strenght of the quadrupole-quadrupole interaction
and the core excitations take the form of
Eq. (\ref{eq:Rigid_rotor_ext_energy}) to accommodate the excited
bands.  The values of $q_{0}^{\rm{band1} \rightarrow \rm{band2}}$,
$E_{K}$ and ${\cal I}$ were calculated as described in the previous
section.

The Hamiltonian matrix for the odd nuclei, to which the theory gives
rise, and which now includes the excited bands, is decomposed into two
parts (in the same fashion as in the previous work) one which is
antisymmetric with respect to particle hole conjugation (for which
physical and unphysical solutions clearly separate into solutions with 
positive and negative energies, respectively) and the other of
which is symmetric. First we diagonalize the antisymmetric part
and identify the physical solutions (those with positive energy).
Then the symmetric part is turned on ``slowly'' and at every step the
physical solutions were identified by checking the eigenvalues of 
a projection
operator which is built from already identified physical wavefunctions
of the previous step.  As soon as the steps of the iteration are small
enough to guarantee that the wavefunctions do not change rapidly
between two steps, this procedure works very well, as it did in our previous 
work. This is because the physical solutions have expectation values close to unity
and  the unphysical ones values close to zero.

\begin{figure}[htbp]
  \begin{center}
    \leavevmode
   \centerline{\hbox{\psfig{figure=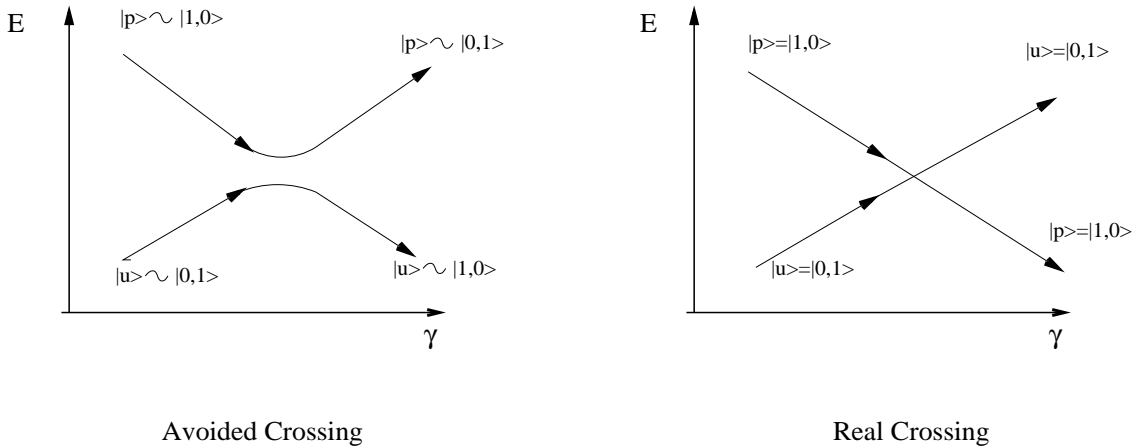,width=15.0cm}} }
  \end{center}
   \caption{ \label{fig:crossing}
Schematic representation of a real crossing (on the right)  ~~~
and an avoided crossing (left). }
\end{figure}

Nevertheless, it is known that for the case that two states with the
same angular momentum come close to each other, they repel and never
really cross (thus the name ``avoided crossing'') (see
Fig.~\ref{fig:crossing}).  Furthermore it is known from sufficiently
general model studies that the wavefunction for each of the bands
changes rapidly in the neighborhood of the crossing and the two end up
inter-changing their character after the crossing.  For example in
Fig.~\ref{fig:crossing} before the crossing the states are almost
equal to the uncoupled states $|1,0 \rangle$ and $|0,1 \rangle$. After
the crossing the state that corresponded to $|1,0 \rangle$, is now
$|0,1 \rangle$ and the state that corresponded to $|0,1 \rangle$ is
$|1,0 \rangle$.

Because the excited bands have band-head energies of about 1~MeV the
possibility of two states coming close to each other is much higher in
the present case than in our previous work.  In order to make sure
that the procedure described above (the projection operator method)
works, we have to make the steps extremely small; as a result the
numerical procedure becomes extremely slow.  To avoid this slowdown,
we have developed a special stratagem. At two successive steps, the
program checks if there was any kind of crossing by inspecting the
differences of eigenvalues of any two eigenstates, identified as
physical or unphysical according to the standard method. If the sign
of the difference changes between two steps, then a crossing has
occurred. (In the case that a crossing is detected, the projection
operator would identify a physical state as unphysical since, as
explained above, their wavefunctions have been interchanged.) Since we
know that every time there is a crossing it is an avoided crossing,
when a crossing is detected we classify a state as physical which
otherwise would have been identified as unphysical from the projection
operator method. Finally the new projection operator will be built
from the new physical wavefunctions. Therefore in the subsequent
steps, we can continue with the projection operator method.

Another technical problem is the classification of states into different bands. 
This was done the same way as in our previous work,
but we include a brief and hopefully clear explanation of the procedure.

Recalling that our formalism remains rotationally invariant even after
omitting the excitation spectrum of the even neighbors, we start in
this limit with a $J=\frac{1}{2}$ submatrix calculation.  The distinct
solutions are identified as $K=\frac{1}{2}$ bands. These levels 
reappear for all higher J calculations at the {\em same} energy. Thus for $J=\frac{3}{2}$,
additional solutions are identified as $K=\frac{3}{2}$ bands, etc.

\begin{figure}[htbp]
  \begin{center}
    \leavevmode
   \centerline{\hbox{\psfig{figure=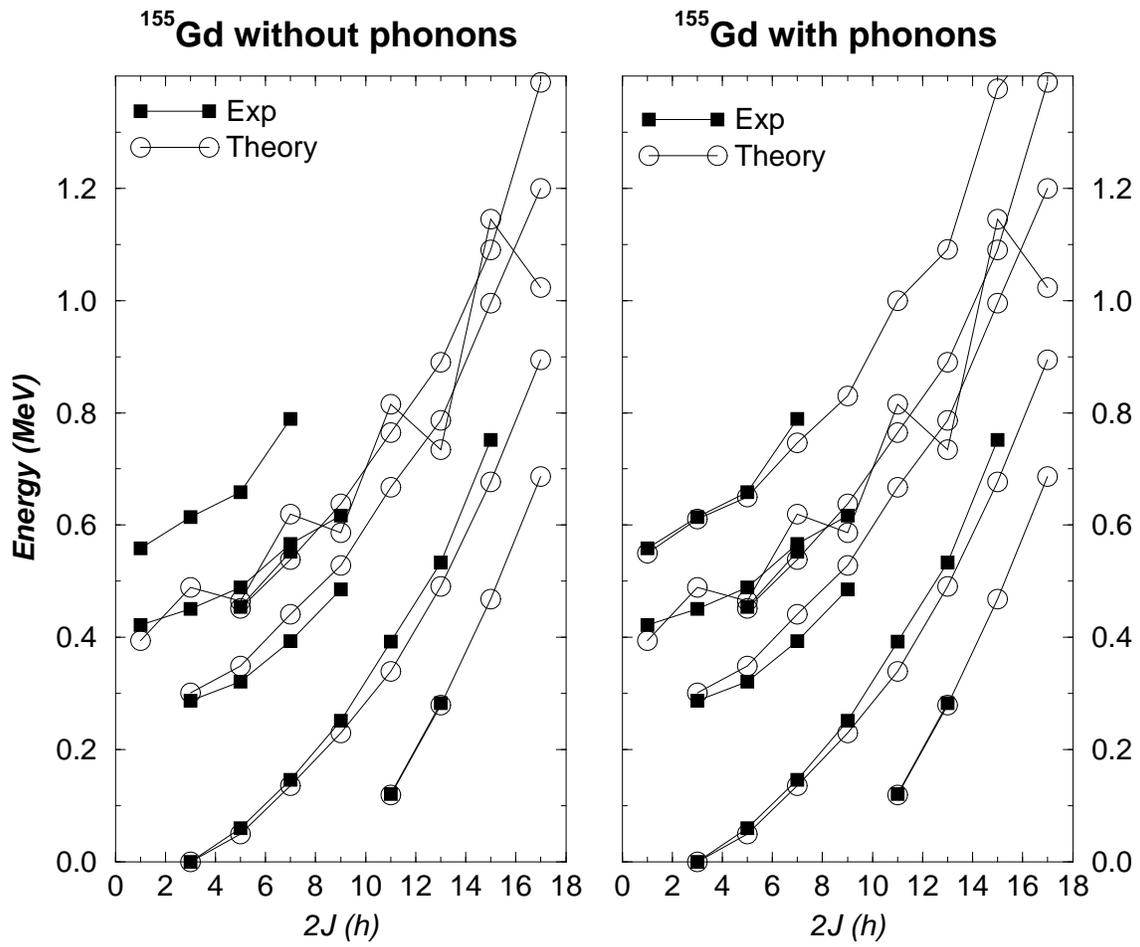,width=15.0cm}} }
  \end{center}
  \caption{  \label{fig:Gd155}
 Energy levels for $^{155}$Gd}
\end{figure}

\begin{figure}[htbp]
  \begin{center}
    \leavevmode
   \centerline{\hbox{\psfig{figure=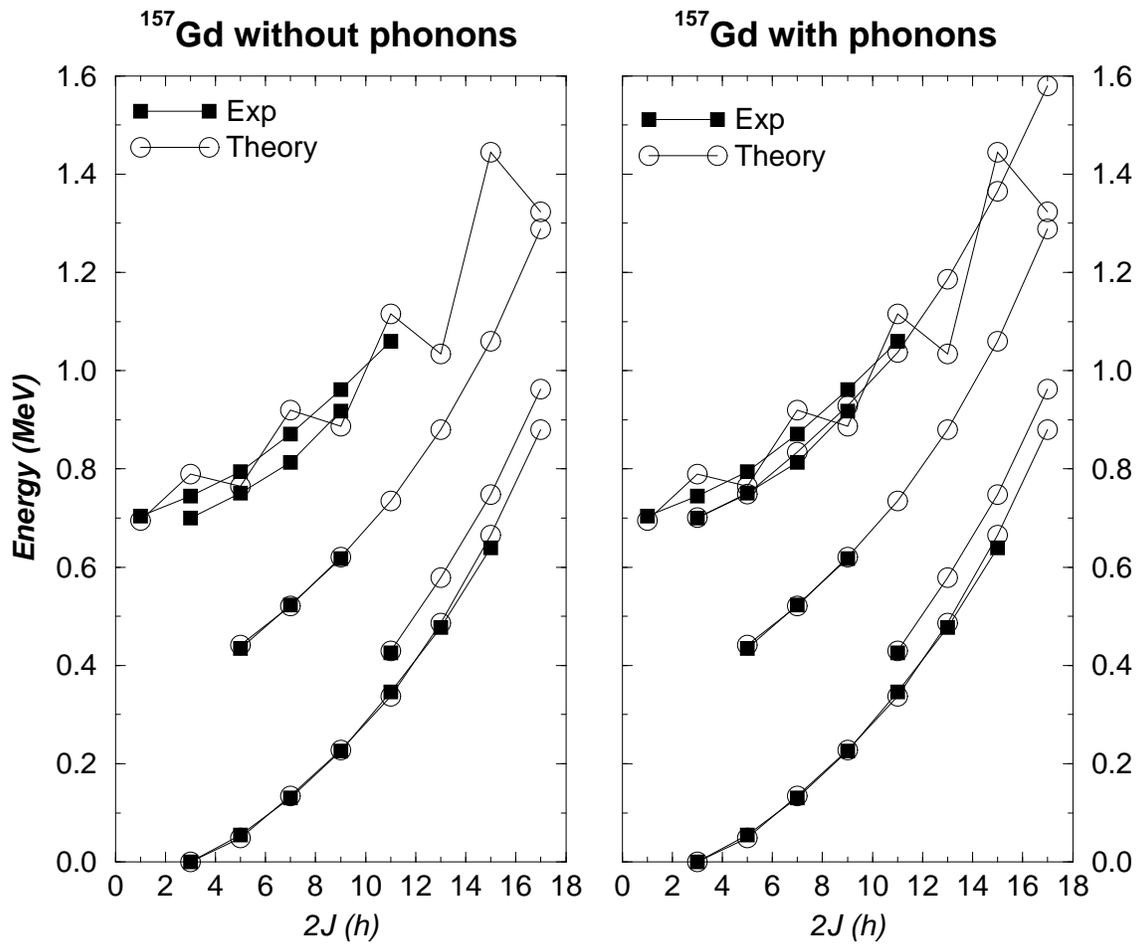,width=15.0cm}} }
  \end{center}
 \caption{  \label{fig:Gd157}
 Energy levels for $^{157}$Gd }
\end{figure}

\begin{figure}[htbp]
  \begin{center}
    \leavevmode
   \centerline{\hbox{\psfig{figure=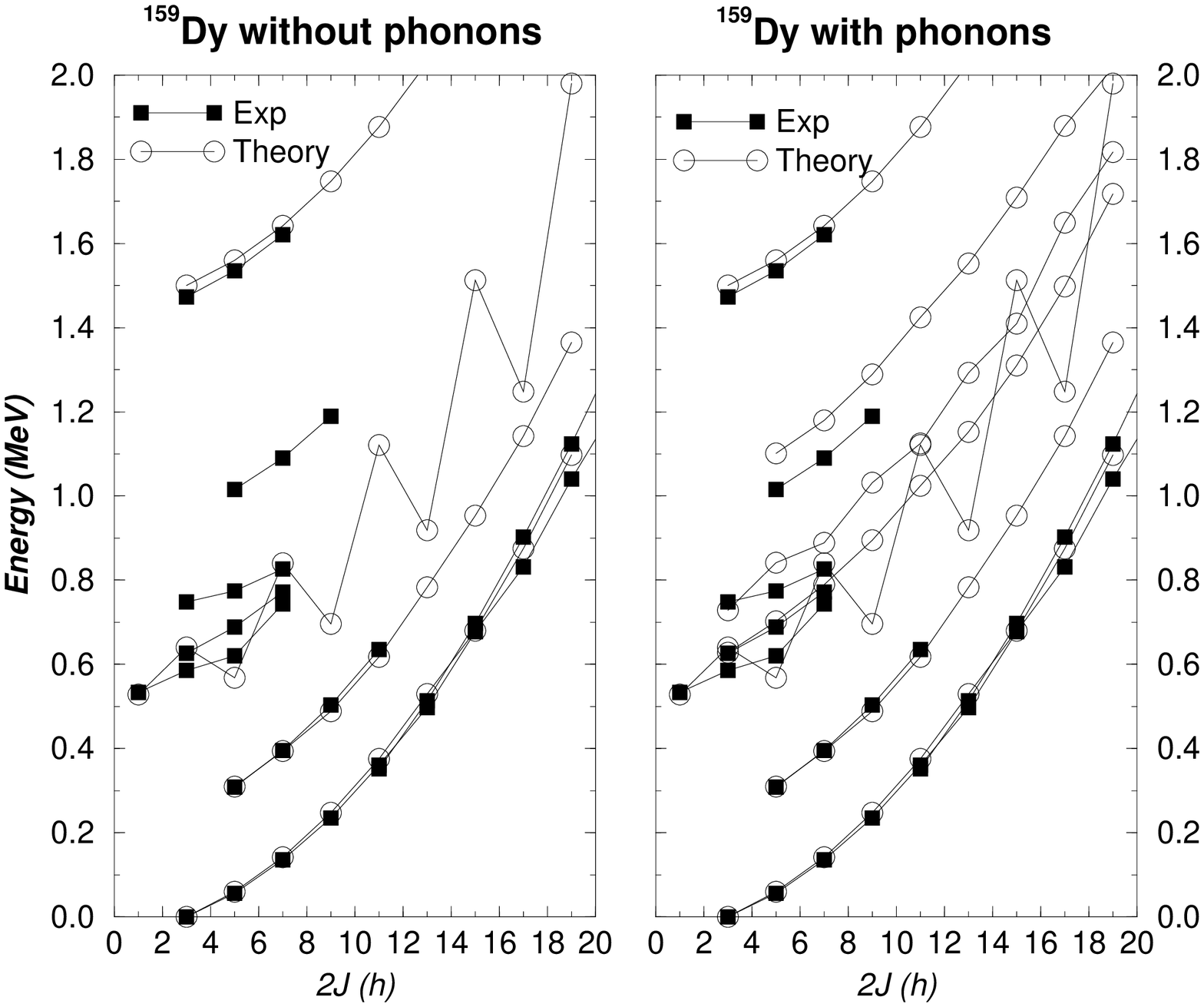,width=15.0cm}} }
  \end{center}
 \caption{   \label{fig:Dy159}
Energy levels for $^{159}$Dy }
\end{figure}

In Figs.~\ref{fig:Gd155},\ref{fig:Gd157} and \ref{fig:Dy159} we show
results of the calculations for three odd nuclei $^{155}$Gd,
$^{157}$Gd and $^{159}$Dy.  In the case of $^{155}$Gd a $K=1/2$ band
at about 0.6~MeV was not reproduced in the calculations without
excited bands, refered to as phonons here.  When phonons were included
in the calculations the missing band was reproduced at the right
band-head energy and almost right band structure. The structure of all
other bands and their relative band-head energies have not changed
more than one percent. This is because the off-diagonal elements of
the Hamiltonian are very small (value of the inter-band transitions)
compared to the diagonal elements. The strength of the quadrupole
force was treated as a free parameter and was fitted to experimental
band-head energies. The best fit was achieved for $\kappa=0.380~{\rm
MeV/fm}^{2}$ which is almost equal to the strength found in the case
of no-phonons ($0.377{\rm MeV/fm}^{2}$). In the case of $^{157}$Gd a
$K=3/2$ band at about 0.7~MeV was not reproduced when phonons were not
included. With the addition of phonons the band was found at the right
band-head energy, and it has the right band structure. As in the case
of $^{155}$Gd the strength of the quadrupole force is almost the same
as for the non-phonon calculations $\kappa=0.401~{\rm MeV/fm}^2$
(compared to $\kappa=0.397~{\rm MeV/fm}^2$). In the last application
we considered $^{159}$Dy. When only the ground state band is allowed,
two experimentally observed bands were not reproduced: $K=3/2$ at
about .7~MeV and $K=5/2$ at about 1~MeV. When a $\beta$ and $\gamma$
excited bands were included in the calculations both bands were
calculated at almost the right band-head energies and have the right
band structure.

\section{Conclusion} 
\label{sec:Pho_con}
Even though the inclusion of the vibration excitation of the even
cores makes the calculations more complicated and numerical solution
longer and more tedious the results supply the previously missing bands.
It thus seems necessary to include the low-lying excited bands of
the neighboring cores in order achieve a good fit to experimental
values. The inclusion of the vibrational bands of the even cores does
not affect the one quasiparticle plus ground-state bands of the odd
nucleus, but some of the low-lying bands are apparently one
quasi-particle plus excited core bands.

\section{Acknowledgement} 
This work was supported in part by U.S. Department of Energy 
under grant number 40264-5-25351

\bibliographystyle{aip}
\bibliography{mybib}

\end{document}